\documentstyle[prl,aps,epsfig,multicol]{revtex}
\newcommand{\beq}{\begin{equation}}
\newcommand{\eeq}{\end{equation}}

\begin{document}

\title{Atom gratings produced by large angle atom beam splitters.}
\author{B. Dubetsky and P. R. Berman}

\address{Michigan Center for Theoretical Physics and Physics Department, University\\
of Michigan, Ann Arbor, MI 48109-1120}

\date{\today}
\maketitle

\begin{abstract}
An asymptotic theory of atom scattering by large amplitude periodic
potentials is developed in the Raman-Nath approximation. The atom grating
profile arising after scattering is evaluated in the Fresnel zone for
triangular, sinusoidal, magneto-optical, and bichromatic field potentials.
It is shown that, owing to the scattering in these potentials, two {\em %
groups} of momentum states are produced rather than two distinct momentum
components. The corresponding spatial density profile is calculated and
found to differ significantly from a pure sinusoid.\vspace{0.03in}

PACS number(s): 03.75, 03.75.D, 39.20, 81.16.T
\end{abstract}

\begin{multicols}{2}

\section{Introduction}

Interaction of atoms with a spatially varying optical field can lead to a
spatial modulation of the atomic density. For example, when a collimated
atomic beam propagating along the $x$ axis passes through a resonant,
standing wave field directed along the $z$ axis, it acquires transverse
momentum components $2n\hbar k$ and $\left( 2n+1\right) \hbar k$ for the
atomic ground and excited states, respectively, where $k$ is the field
propagation constant and $n$ is an integer \ Interference between these
momentum states \cite{c1} results in a periodic modulation of the atom
density with period $d=\lambda /2,$ where $\lambda =2\pi /k$. This splitting
of the atomic wave function (resonant Kapitza-Dirac effect \cite{c2}) and
the density modulation have been observed in a beam of Na \cite{c3} and Ar 
\cite{c4} atoms, respectively. Atom gratings can be deposited on substrates
and have potential applications as diffraction gratings for soft x-rays.

The grating density profile $\rho \left( z\right) $ produced by this
technique contains an infinite number of Fourier harmonics having periods $%
\lambda /2n$ and amplitudes $\rho _{n}.$ For the grating profile to be
approximately sinusoidal, then at most a few $\rho _{n}$ contribute in the
Fourier sum$.$ On the other hand, the ''sharp'' grating that can be produced
by atom focusing using strong standing wave fields \cite{c5,c6} consists of
a large number of Fourier components, 
\begin{equation}
n\sim \lambda /w\gg 1,  \label{2}
\end{equation}
where $w\ll \lambda $ is the spot size of the focussed atoms$.$ Sharp
gratings have important lithographic applications, but are not well suited
for use as elements to diffract x-rays. For x-ray scattering, one finds that
each density component $\rho _{n}$ leads to scattering at angle 
\begin{equation}
\theta _{n}=\pm n\lambda _{x}/d,  \label{3}
\end{equation}
where $\lambda _{x}\ll \lambda $ is the x-ray wavelength. If one monitors
scattering at a specific angle, most of the scattered energy does not
contribute to the signal. As a consequence, it is most efficient to scatter
x-rays from a sinusoidal atom grating, since only one Fourier component is
nonvanishing $\left[ n=1\text{ in Eq. (\ref{3})}\right] .$ To increase the
angle and resolution of scattering one needs a high order sinusoidal grating
having period 
\begin{equation}
d=\lambda /2\bar{n},  \label{4}
\end{equation}
where $\bar{n}>1$. Several techniques have been proposed to produce such
high order gratings. One can distinguished two groups of methods, harmonic
suppression (HS) and large angle beam splitters (LABS) . In HS, one arrives
at a periodicity (\ref{4}) as a result of suppressing all Fourier components
in the atom wave function that are not a multiple of $\bar{n}.$ Included in
the HS group are methods that employ photon echoes \cite{c8}, multicolor
fields \cite{c9}, high order Bragg scattering \cite{c10}, and Fourier optics
techniques \cite{J}, methods which are summarized below.

An appropriate echo geometry for HS involves atoms that pass through two
standing wave fields, spatially separated by a distance $L$ (or atoms that
are subjected to two temporally separated standing wave field pulses). At
distances $L^{\prime }=\frac{n^{\prime }}{\bar{n}}L$ ($n^{\prime }>\bar{n}$,
where $n^{\prime }$ and $\bar{n}$ are integers having no common factors),
the atomic density contains only those Fourier components having $n$ that is
a multiple of $\bar{n}.$ The other components are suppressed if the atom
beam has a transverse velocity width $\delta \theta >\lambda /L$ \cite{c8}.
A $\frac{\lambda }{4}-$period grating $\left( \bar{n}=2\right) $ has been
observed using this method on $^{85}$Rb atoms cooled in a MOT and subjected
to the 2 time-separated standing wave field pulses \cite{c11}. In the
multicolor field technique \cite{c9}, atoms interact during time $\tau $
with a traveling wave field having frequency $\Omega $ and two
counterpropagating traveling wave fields having frequencies $\Omega +\delta
_{1}$ and $\Omega +\delta _{2}.$ If $\left| \delta _{j}\tau \right| \gg 1$
and the detunings are chosen such as $n_{1}\delta _{1}+n_{2}\delta _{2}=0,$
then a grating having period $\lambda /2\bar{n}$ is produced, with $\bar{n}%
=n_{1}+n_{2}.$ In contrast to the echo-technique, the multicolor field
method allows one to obtain near-unity contrast for arbitrary $\bar{n}.$ For
Bragg scattering, a proper choice of the angle between the atomic beam and
the field propagation direction or, alternatively, a proper choice of
atom-field detunings \cite{c12}, leads to situations where low spatial
harmonics are suppressed. A $\frac{\lambda }{6}-$period grating has been
obtained using this technique \cite{c10}. In the Fourier optics technique
one focuses atoms transmitted through a standing wave field using a
long-focus lens. Different harmonics of the atomic wave function are focused
at the different positions. Using spatial filtering, one can suppress all
harmonics except two, whose interference leads to the grating of the
desirable period. In this way one achieves 100\% contrast pure sinusoidal
gratings, but the amplitude of the gratings decreases with increasing
grating order.

An ideal large angle beam splitter (LABS) splits an initial state having
transverse momentum $p=0$ into two states having momenta 
\begin{equation}
p=\pm \bar{n}\hbar k,  \label{41}
\end{equation}
where $\bar{n}\gg 1$. The interference between these states leads to a
spatial density 
\begin{equation}
\rho \left( z\right) =1+\cos \left( 2\bar{n}kz\right)   \label{5}
\end{equation}
having period $\lambda /2\bar{n}$. The prototype LABS consists of a
V-potential 
\begin{equation}
U\left( z,t\right) =f\left( t\right) \hbar k\bar{n}\left| z\right| ,
\label{6}
\end{equation}
off of which atoms scatter. The function $f(t),$ normalized such that $%
\int_{-\tau /2}^{\tau /2}f(t)dt=1$, describes the potential experienced by
an atom in its rest-frame [$t=x/u,$ where $u$ is the projection of the atom
velocity along the beam propagation direction ($x$ axis) and $\tau $ is the
pulse duration in the atomic rest frame]. In the Raman-Nath approximation 
\begin{equation}
\omega _{\bar{n}k}\tau \ll 1,  \label{61}
\end{equation}
where 
\[
\omega _{q}=\hbar q^{2}/2m
\]
is a recoil frequency and $m$ is the atomic mass, one can show that, at a
time $t$ following the atom field interaction and for $\bar{n}\gg \left(
\omega _{k}t\right) ^{-1/2}$, an atom grating is produced having density 
\begin{eqnarray}
\rho \left( z\right)  &\sim &2\left[ 1+\cos \left( 2\bar{n}kz\right) \right] 
\text{ \ \ for }\left| z\right| <\bar{n}\frac{\hbar k}{m}t;  \nonumber \\
&\sim &1\text{ \ \ for }\left| z\right| >\bar{n}\frac{\hbar k}{m}t.
\label{611}
\end{eqnarray}

In the past several years, a number of methods have been proposed for using
counterpropagating optical fields to create potentials that simulate a
V-potential (\ref{6}). Atom scattering off these fields involves multiphoton
processes that result in a transfer of momentum between the
counterpropagating field modes. The effective potential $U\left( z,t\right) $
corresponding to these processes is a function of bi-linear products of the
counterpropagating field amplitudes. As a consequence, $U\left( z,t\right) $
has period $\lambda /2$ as a function of $z.$ One might expect that a
so-called {\em triangular potential}, a periodic potential coinciding with a
V-potential within a period,

\begin{eqnarray}
U\left( z,t\right) &=&f\left( t\right) \hbar k\bar{n}\left\{ 
\begin{tabular}{l}
-$z,$ for $\left| z\right| <\lambda /8$ \\ 
$z-\lambda /4,$ for $\left| z-\lambda /4\right| <\lambda /8$%
\end{tabular}
\right. ,  \nonumber \\
U\left( z+\lambda /2,t\right) &=&U\left( z,t\right) ,  \label{62}
\end{eqnarray}
should produce gratings similar to the sinusoidal grating (\ref{611})
produced by the V-potential. Based on this idea several schemes for LABS
have been proposed.

The simplest example of such a beam splitter is the strong standing wave
field first considered in \cite{c2}. After adiabatic elimination of the
internal degrees of freedom, one finds a sinusoidal potential having period $%
\frac{\lambda }{2}$. Asymptotically, for a large potential amplitude, the
main contribution to the scattering arises from the area around points $z_{0}
$ where the potential slope is maximum; in this region the potential is
approximately linear. The momentum an atom acquires after scattering near
such points is $p\sim F\tau ,$ where $F\sim -\left( \partial U/\partial
z\right) _{z=z_{0}}$ is the classical force acting on atom.

An alternative method which produces a potential that more closely
approximates (\ref{62}) is the {\em magneto-optical field geometry }\cite
{c13}, in which resonant cross-polarized counterpropagating fields drive a
ground-to-excited state transition, $J=0\rightarrow J=1,$ where $J$ is the
atomic angular momentum. A magnetic field directed along the $z$ axis that
splits the excited state Zeeman sublevels is also applied. A regime has been
found where the magneto-optical potential is extremely close to the
triangular potential (\ref{62}). The nonlinearity near $z=z_{0}$ is small
compared with that associated with the standing wave potential.
Magneto-optical forces have been used to deflect an atomic beam \cite{c13}
and to split an atomic beam into two momenta components localized near
values $p=\pm 21\hbar k$ [$\bar{n}=21$ in Eq. (\ref{41})] \cite{c14}. Atom
scattering from a magneto-optical potential is considered in \cite
{c15,c16,p3}, while the force acting on atoms is calculated in \cite{c17}.
Instead of using a magnetic field one can detune standing waves
symmetrically with respect to the atomic transition frequency \cite{p1}.

Another way of creating a potential similar to the triangular one is to use
a bichromatic field \cite{c171}, in which atoms scatter from two standing
wave fields, shifted from one another with respect to their frequencies, and
spatial and temporal phases. Forces resulting from this field have been
observed in \cite{c172,c173}. A remarkable property of these forces (large
magnitude in comparison with radiation forces, insensitivity to atomic
velocity) allows for fast atomic beam deceleration \cite{c174} and for large
angle deflections \cite{c175}. The triangular potential associated with a
bichromatic field was predicted in \cite{c176}, and scattering off this
field has been observed \cite{p2} and studied for two-level \cite{Walls,p2}
and three level atoms \cite{c177,c178,c179,c180}. Moreover, it has been
shown \cite{c181} that atoms can be scattered with unit probability with a
given change of momentum if a ''counterintuitive pulse sequence'' is used.

Despite the similarity between the triangular potential (\ref{62}) and
V-potential (\ref{6}), there is an important difference. Owing to the
intrinsic $\lambda /2$ periodicity, atoms scatter in the potential (\ref{62}%
) not only into states $p=\pm \bar{n}\hbar k$ but also into states having $%
p=\pm \left( \bar{n}\pm 2\right) \hbar k,$ $\pm \left( \bar{n}\pm 4\right)
\hbar k$ $\ldots $. Thus, at best, the potential (\ref{62}) can produce two 
{\em groups} of momenta centered at $p=\pm \bar{n}\hbar k.$ As such the
resulting pattern deviates from that of a true V-potential. The additional
momentum states produce deviations from an ideal sinusoidal atom grating (%
\ref{5}) by producing additional Fourier components, 
\begin{equation}
\rho _{\nu }\left( z\right) =a_{\nu }\cos \left[ 2\left( \bar{n}+\nu \right)
kz\right] ,  \label{7}
\end{equation}
where $\alpha _{\nu }\sim 1$ for $\nu \sim \pm \Delta n/2$ and $\Delta n\ll 
\bar{n}$ is the range of components created by the potential. Therefore,
LABS produces atom gratings that are not pure, high order gratings; their
overall periodicity is still $\lambda /2.$ Gratings arising in HS also
contain extra terms (\ref{7}), but one is guaranteed that $a_{\nu }\ll a_{0}$
if $\nu $ is not a multiple of $\bar{n}.$ As a result, atom gratings
produced by HS closely approximate high order $\frac{\lambda }{2\bar{n}}$%
-period gratings, having an envelope that oscillates slightly with period $%
\lambda /2$ around a constant value$.$

Scattering into a range of momentum components occurs even for the
triangular potential. For potentials which approximate the triangular
potential (standing wave, magneto-optical scheme, bichromatic field),
additional features arise owing to the nonlinearities in these potentials.
It turns out that the width $\Delta n$ for these potentials increases with
increasing $\bar{n}.$ If one tries to decrease the grating period by
increasing $\bar{n}$, there is a corresponding deterioration of the degree
to which the grating profile approximates a pure sinusoidal grating.

In this article we provide a detailed calculation of the spatial gratings
produced by LABS, using analytical asymptotic and\ numerical calculations.
We consider only the Fresnel regime when all scattered atomic beams overlap
with one another. An asymptotic expression for the atomic wave function in
momentum space is derived in the next Section and the atomic density profile
is calculated in Sec. 3. Results are summarized and discussed in Sec. 4. In
the Appendix, we consider atom gratings produced by the triangular potential
(\ref{62}).

\section{Atom wave function}

In this section, we calculate the atomic response to a standing wave field
pulse, a magneto-optical field pulse and a bichromatic field pulse, all in
the Raman-Nath approximation. For an atomic beam, the pulsed interaction
occurs in the atom rest frame.

\subsection{Standing wave.}

When a two-level atom interacts with a standing wave field pulse having
electric field, 
\begin{equation}
{\bf E}\left( t\right) ={\bf \hat{y}}E_{0}\left( t\right) \cos \left(
kz\right) e^{-i\omega t}+c.c.,  \label{8}
\end{equation}
having (real) amplitude $E_{0}\left( t\right) $, propagation constant $k$,
duration $\tau $, and frequency $\omega $, the atomic wave function ${\bf %
\Psi }=\left( 
\begin{tabular}{c}
$\psi $ \\ 
$\psi _{e}$%
\end{tabular}
\right) $ (in the interaction representation) evolves as 
\end{multicols}
\begin{equation}
i{\bf \dot{\Psi}}=\left( 
\begin{tabular}{cc}
0 & $\chi \left( t\right) \cos \left( kz\right) \exp \left( i\Delta t\right) 
$ \\ 
$\chi \left( t\right) \cos \left( kz\right) \exp \left( -i\Delta t\right) $
& 0
\end{tabular}
\right) {\bf \Psi },  \label{9}
\end{equation}
where $\psi $ and $\psi _{e}$ are ground ($g$) and excited ($e$) internal
states wave functions, $\Delta =\omega -\omega _{0}$ is the atom-field
detuning, $\chi \left( t\right) $=$\mu E_{0}\left( t\right) /\hbar $ is a
Rabi frequency, $\omega _{0}$ is the ground to excited state transition
frequency, and $\mu $ is a (real) dipole moment operator matrix element
associated with the $e\rightarrow g$ transition$.$ In Eq. (\ref{9}) we
neglect a term, $\left( \hat{p}^{2}/2m\hbar \right) {\bf \Psi }$, ($\hat{p}$
is the atomic center-of-mass momentum operator), which is valid in the
Raman-Nath approximation (\ref{61}). For an adiabatic pulse 
\begin{equation}
\dot{\chi}/\chi \ll \max \left\{ \left| \Delta \right| ,\left| \chi \right|
\right\} ,  \label{10}
\end{equation}
it is convenient to use a semiclassical dressed state approach \cite{c18}.
If the wave function before the pulse arrives is given by ${\bf \Psi }%
_{-}=\left( 
\begin{tabular}{c}
$\psi _{-}\left( z\right) $ \\ 
$0$%
\end{tabular}
\right) ,$ then, immediately following the pulse, the atom returns
adiabatically to its ground state and the wave function is ${\bf \Psi }%
_{+}=\left( 
\begin{tabular}{c}
$\psi _{+}\left( z\right) $ \\ 
$0$%
\end{tabular}
\right) ,$ where 
\begin{mathletters}
\label{11}
\begin{eqnarray}
\psi _{+}\left( z\right) &=&\eta \left( z\right) \psi _{-}\left( z\right) ,
\label{11a} \\
\eta \left( z\right) &=&\exp \left[ -i\theta \left( z\right) \right] ,
\label{11b} \\
\theta \left( z\right) &=&\int_{-\infty }^{\infty }dt\left\{ \left[ \frac{%
\Delta ^{2}}{4}+\frac{\chi ^{2}\left( t\right) }{2}\left( 1+\cos \left(
2kz\right) \right) \right] ^{1/2}-\frac{\Delta }{2}\right\} ,  \label{11c}
\end{eqnarray}
$\eta \left( z\right) $ is a transmission function and $\theta \left(
z\right) $ is the spatially inhomogeneous pulse area. To simplify the
calculations, it is assumed that the pulses considered in this paper have
leading and trailing edges of duration $\tau _{r}$, and that the pulse Rabi
frequency is constant and equal to $\chi $ between the leading and trailing
edges. Adiabaticity is guaranteed if $\left| \Delta \right| \tau _{r}\gg 1.$
Moreover, if $\tau _{r}/\tau \ll 1,$ the contribution to the phase $\theta
\left( z\right) $ from the tails is smaller by a factor of order $\tau
_{r}/\tau $ from that of the central region and modifies only slightly the
points of stationary phase to be calculated below. As such we approximate $%
\theta \left( z\right) $ as

\end{mathletters}
\begin{mathletters}
\label{12}
\begin{eqnarray}
\theta \left( z\right) &=&\theta w\left( z\right) ,  \label{12a} \\
w\left( z\right) &=&\left\{ 1+\xi \left[ 1+\cos \left( 2kz\right) \right]
\right\} ^{1/2}  \label{12b}
\end{eqnarray}
where 
\end{mathletters}
\[
\theta =\Delta \tau /2 
\]
and 
\[
\xi =2\chi ^{2}/\Delta ^{2}. 
\]
If initially an atom is in the zero momentum state, then the Fourier
components 
\begin{equation}
\psi _{n}=\int_{0}^{\lambda /2}\frac{dz}{\lambda /2}\exp \left[
-2inkz-i\theta (z)\right]  \label{13}
\end{equation}
determine the amplitudes for scattering into momentum states $p=2n\hbar k,$
which we refer to as scattering into state $n.$

In the asymptotic limit 
\begin{equation}
\theta _{\max }-\theta _{\min }\gg 1,  \label{14}
\end{equation}
the main contribution to the integral (\ref{13}) arises from points of
stationary phase, $z=z_{0}.$ These contributions are maximal if the second
derivative of the phase also vanishes at $z=z_{0}$. This means that the
dominant contribution to scattering into state $n$ occurs for $n\approx 
\tilde{n}_{0},$ where $\tilde{n}_{0}$ is determined by the equations 
\begin{mathletters}
\label{15}
\begin{eqnarray}
\tilde{n}_{0} &=&-\theta ^{\prime }\left( z_{0}\right) /2k,  \label{15a} \\
\theta ^{\prime \prime }\left( z_{0}\right) &=&0.  \label{15b}
\end{eqnarray}
Expanding the phase in (\ref{13}) to third order in $\left( z-z_{0}\right) ,$
one finds that $\psi _{n}$ is given by the asymptotic expression, 
\end{mathletters}
\begin{equation}
\psi _{n}\sim 2k\left[ 2/\left| \theta ^{\prime \prime \prime }\left(
z_{0}\right) \right| \right] ^{1/3}\exp \left[ -2inkz_{0}-i\theta \left(
z_{0}\right) \right] Ai\left\{ \Delta n2k\left[ 2/\theta ^{\prime \prime
\prime }\left( z_{0}\right) \right] ^{1/3}\right\} ,  \label{16}
\end{equation}
\begin{multicols}{2}
where 
\[
\Delta n=n-\tilde{n}_{0} 
\]
and $Ai\left( x\right) $ is an Airy function. For the field parameters given
in (\ref{12}), one has 
\begin{mathletters}
\label{17}
\begin{eqnarray}
\tilde{n}_{0} &=&\pm n_{0};  \label{17a} \\
n_{0} &=&\theta \left\{ \left[ 1+\xi -\left( 1+2\xi \right) ^{1/2}\right]
/2\right\} ^{1/2}, \\
\theta ^{\prime \prime \prime } &=&\pm 8k^{3}n_{0},  \label{17b} \\
z_{0} &=&\pm \frac{\lambda }{4\pi }\cos ^{-1}\left\{ \left[ \left( 1+2\xi
\right) -1-\xi \right] /\xi \right\} ,  \label{17c} \\
\theta \left( z_{0}\right) &=&\theta \left( 1+2\xi \right) ^{1/4},
\label{17d}
\end{eqnarray}
such that 
\end{mathletters}
\begin{equation}
\left| \psi _{n}\right| ^{2}\sim \left( 2/n_{0}\right) ^{2/3}Ai^{2}\left[
\Delta n\left( 2/n_{0}\right) ^{1/3}\right] .  \label{18}
\end{equation}
The distribution (\ref{18}) is plotted in Fig. \ref{f1} and has width
(full-width at half maximum) 
\begin{equation}
\Delta n_{sw}\approx 1.29n_{0}^{1/3}.  \label{19}
\end{equation}
The result (\ref{18}) differs from that obtained in reference \cite{c2}.

Using an asymptotic expression for the Airy function for large negative
arguments one finds that for $\Delta n<0$ and $\left| \Delta n\right| \gg 1$
the distribution oscillates as a function of $\Delta n.$ Averaging over
these oscillations one finds 
\begin{equation}
\left| \psi _{n}\right| ^{2}\sim \pi ^{-1}\left( 2n_{0}\Delta n\right)
^{-1/2}.  \label{191}
\end{equation}
which coincides with the momentum distribution associated with classical
particles scattered by a standing wave potential \cite{G}.

In the case of a far detuned field $\left( \xi \ll 1\right) ,$ when 
\begin{mathletters}
\label{17}
\begin{eqnarray}
\theta \left( z\right) &=&\theta \left\{ 1+\frac{\xi }{2}\left[ 1+\cos
\left( 2kz\right) \right] \right\} ,  \label{21a} \\
n_{0} &=&\theta \xi /2,  \label{21b}
\end{eqnarray}
one finds from Eq. (\ref{13}) that the exact expression for the Fourier
harmonics is 
\end{mathletters}
\begin{equation}
\psi _{n}=J_{n}\left( \theta \xi /2\right) \exp \left[ -in\pi /2-i\theta
(1+\xi /2)\right] ,  \label{22}
\end{equation}
where $J_{n}\left( x\right) $ is a Bessel function. The asymptotic result (%
\ref{16}) follows directly from Eq. (\ref{22}) using a well-known asymptotic
expression for large order Bessel functions \cite{c19}. One can consider Eq.
(\ref{16}) as a generalization of the asymptotic expression for Bessel
functions in the limit that $n\gg 1$ for a function having arbitrarily large
periodical phase.

\subsection{Magneto-optical potential}

This regime of scattering arises \cite{c13}, when atoms interact with an
optical field 
\begin{equation}
{\bf E(}z,t)=E_{0}\left( t\right) \left( {\bf \hat{y}}e^{ikz}+{\bf \hat{x}}%
e^{-ikz}\right) e^{-i\omega t}+c.c.  \label{23}
\end{equation}
and a static magnetic field along ${\bf \hat{z}}.$ If the optical field
drives a $J=0\rightarrow J=1$ ground-to-excited state transition, then the
wave function ${\bf \Psi }^{\prime }{\bf =}\left( 
\begin{tabular}{l}
$\psi ^{\prime }$ \\ 
$\psi _{-e}$ \\ 
$\psi _{e}$%
\end{tabular}
\right) ,$ ($\psi ^{\prime }=\psi e^{i\Delta t},$ $\psi _{-e},$ $\psi _{e}$
are amplitudes for the ground state, $m_{j}=-1$ excited state$,$ and $%
m_{j}=1 $ excited state, respectively) evolves as 
\begin{equation}
i{\bf \dot{\Psi}}^{\prime }{\bf =U\Psi }^{\prime }{\bf ,}  \label{24}
\end{equation}
where the interaction Hamiltonian is 
\begin{equation}
\hbar {\bf U}=\hbar \left( 
\begin{tabular}{ccc}
$\Delta $ & $G_{-}$ & $G_{+}$ \\ 
$G_{-}^{\ast }$ & -$\omega _{L}$ & $0$ \\ 
$G_{+}^{\ast }$ & $0$ & $\omega _{L}$%
\end{tabular}
\right) ,  \label{25}
\end{equation}
$\pm \omega _{L}$ are Zeeman shifts of the states $\pm e,$ 
\[
G_{\pm }=\mp 2^{-1/2}\omega _{R}e^{\pm i\pi /4}\cos \left( kz\mp \pi
/4\right) , 
\]
$\omega _{R}=2\times 3^{-1/2}E_{0}^{\ast }\left( t\right) \mu /\hbar $ is a
Rabi frequency, and $\mu $ is a reduced matrix element of the dipole moment
operator. To develop an asymptotic solution of Eq. (\ref{24}), one must
first solve the characteristic equation 
\begin{equation}
\left| U-\lambda I\right| =0.  \label{26}
\end{equation}
Solutions can be found in Refs. \cite{c15,c16,c17}. Let us denote by $%
\lambda \left( z,t\right) $ the root of Eq. (\ref{26}) which equals $\Delta $
when the field vanishes. If initially an atom is in its ground state and if
there is no level crossing during the interaction, then after the pulse the
atom returns to its ground state. The ground state wave function $\psi $
following the pulse is given by equations analogous to Eqs. (\ref{11}) with
a pulse area defined by 
\begin{equation}
\theta \left( z\right) =\int_{-\infty }^{\infty }dt\left[ \lambda \left(
z,t\right) -\Delta \right] .  \label{27}
\end{equation}

A potential that closely resembles the triangular potential (\ref{6})
emerges if one chooses a rectangular pulse having \cite{c13} 
\begin{equation}
\omega _{R}=2\omega _{L}\text{ and }\Delta =0  \label{271}
\end{equation}
(although $\Delta =0$, adiabaticity is still maintained if $\omega _{L}\tau
_{r}\gg 1$). For these values, we find from Eqs. (\ref{25}) and (\ref{27})
that 

\begin{figure}
\begin{minipage}{0.97\linewidth}
\begin{center}
\epsfxsize=.97\linewidth \epsfysize=2.01\linewidth \epsfbox{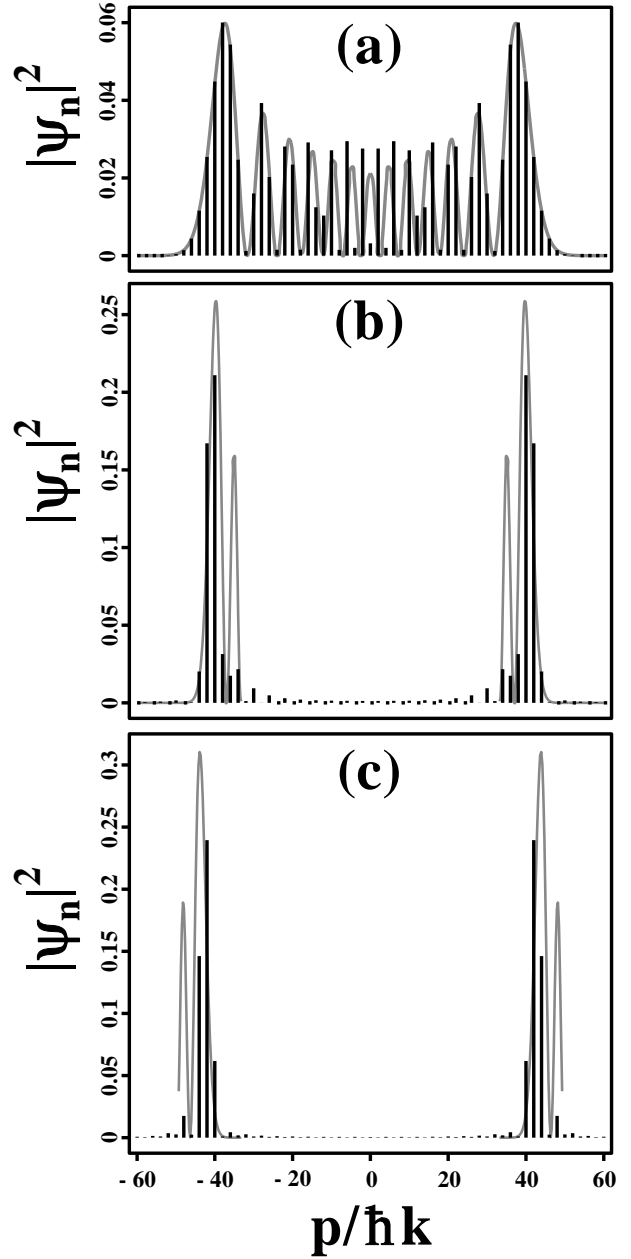}
\end{center}
\end{minipage}
\begin{minipage}{0.99\linewidth} \caption{Probabilities $\left| \protect\psi _{n}\right| ^{2}$ for atom
scattering by LABS into momentum states $p=n\hbar k$ located at $n\approx
\pm n_{0}$ for $n_{0}=20\protect\pi /3,$ corresponding to the modulation
amplitude chosen in \protect\cite{c15} for numerical calculations; $\left(
a\right) $ standing wave field, $\left( b\right) $ magneto-optical field, $%
\left( c\right) $ bichromatic field. Solid lines are the numerical values
and the gray solid lines are asymptotic expressions obtained from Eqs. (\ref
{18}, \ref{30}, \ref{3106}). 
\label{f1}}
\end{minipage}
\end{figure}
\begin{mathletters}
\label{28}
\begin{eqnarray}
\theta \left( z\right)  &=&-\theta w\left( z\right) ,  \label{28a} \\
w\left( z\right)  &=&\sin \left\{ \frac{1}{3}\sin ^{-1}\left[ \sin \left(
2kz\right) \right] \right\} ,  \label{28b}
\end{eqnarray}
where $\theta =2\omega _{L}\tau .$ The function $w\left( z\right) $ has
discontinuous derivatives. In the regions of continuity this function is
given by 
\end{mathletters}
\begin{mathletters}
\label{29}
\begin{eqnarray}
w\left( z\right)  &=&\sin \left( \frac{2}{3}kz\right) ,\text{ for }0<z<\frac{%
\lambda }{8},  \label{29a} \\
&=&\sin \left( \frac{\pi -2kz}{3}\right) ,\text{ for }\frac{\lambda }{8}<z<%
\frac{3}{8}\lambda ,  \label{29b} \\
&=&\sin \left[ \frac{2}{3}\left( kz-\pi \right) \right] ,\text{ for }\frac{3%
}{8}\lambda <z<\frac{\lambda }{2}.  \label{29c}
\end{eqnarray}

The slope of $w\left( z\right) $ has extrema at $z_{0}=0$ and $z_{0}=\lambda
/4,$ associated with splitting of the incident state in two momentum groups
located at $\tilde{n}_{0}=\pm n_{0},$ with 
\end{mathletters}
\begin{equation}
n_{0}=\theta /3.  \label{291}
\end{equation}
From Eq. (\ref{16}) it follows that 
\begin{equation}
\psi _{n}\sim \left( \pm 1\right) ^{n}\left( 18/n_{0}\right) ^{1/3}Ai\left[
\pm \Delta n\left( 18/n_{0}\right) ^{1/3}\right] .  \label{30}
\end{equation}
Exact and asymptotic values for $\left| \psi _{n}\right| ^{2}$ as a function
of $n$ are shown in Fig. \ref{f1}. One sees that asymptotically, for a given
angle of splitting (i. e. a given value of $n_{0}),$ the net result of
replacing a standing wave field by a magneto-optical field is to increase
the maximum of the scattering probability by a factor $3^{4/3}$ and decrease
of the distribution width $\Delta n_{mo}$ by a factor $3^{2/3}.$ In spite of
this, the width of distribution is still an increasing function of the angle
of scattering, 
\begin{equation}
\Delta n_{mo}=0.62n_{0}^{1/3}  \label{31}
\end{equation}

\subsection{Bichromatic field}

\begin{figure}
\begin{minipage}{0.97\linewidth}
\begin{center}
\epsfxsize=.97\linewidth \epsfysize=.96\linewidth \epsfbox{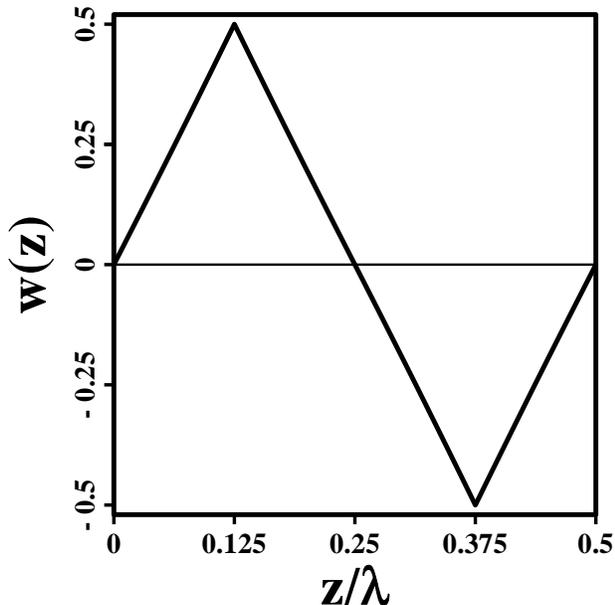}
\end{center}
\end{minipage}
\begin{minipage}{0.99\linewidth} \caption{One period of the\ normalized pulse area $w\left( z\right) =\protect%
\theta \left( z\right) /\protect\theta $ for scattering by a bichromatic
field. 
\label{f2}}
\end{minipage}
\end{figure}

Finally, we consider the interaction of a two-level atom with a quantized
field composed of two standing wave fields having frequencies $\omega
_{0}\pm \Delta ,$ the same initial number of photons $\nu \gg 1,$ and
shifted with respect to one another by a quarter of a wavelength. The
Hamiltonian can be written as \cite{Walls} 
\end{multicols}
\begin{mathletters}
\label{3101}
\begin{eqnarray}
H &=&H_{0}+H_{I},  \label{3101a} \\
H_{0} &=&\hbar \omega _{0}\left( \left| e\right\rangle \left\langle e\right|
+c^{\dagger }c+d^{\dagger }d\right) ,  \label{3101b} \\
H_{I} &=&\hbar \Delta \left( c^{\dagger }c-d^{\dagger }d\right) -\hbar
g\left\{ \cos \left( kz+\pi /4\right) \left[ c^{\dagger }\left|
g\right\rangle \left\langle e\right| +\left| e\right\rangle \left\langle
g\right| c\right] +\cos \left( kz-\pi /4\right) \left[ d^{\dagger }\left|
g\right\rangle \left\langle e\right| +\left| e\right\rangle \left\langle
g\right| d\right] \right\} ,  \label{3101c}
\end{eqnarray}
\end{mathletters}
\begin{multicols}{2}
where $c$ and $d$ are annihilation operators for modes $\omega _{0}\pm
\Delta ,$ $g$ is a coupling constant, and $H_{I}$ is the interaction
Hamiltonian. If initially an atom is in its ground state, then in the
interaction representation the system wave function is given by 
\begin{equation}
{\bf \Psi =}\left( 
\begin{tabular}{c}
$\vdots $ \\ 
$\psi _{g}\left( \nu -1,\nu +1;z,t\right) $ \\ 
$\psi _{e}\left( \nu -1,\nu ;z,t\right) $ \\ 
$\psi \left( z,t\right) $ \\ 
$\psi _{e}\left( \nu ,\nu -1;z,t\right) $ \\ 
$\psi _{g}\left( \nu +1,\nu -1;z,t\right) $ \\ 
$\vdots $%
\end{tabular}
\right) ,  \label{3102}
\end{equation}
where $\psi _{i}\left( m,n;z,t\right) $ is the amplitude for the atom to be
in state $\left| i\right\rangle $ and the field to be in a state having $m$
and $n$ photons in the modes $\omega _{0}\pm $ $\Delta ,$ respectively.
Before the atom field interaction, $\psi \left( z,t\right) \equiv \psi
_{g}\left( \nu ,\nu ;z,t\right) $. This wave function evolves as 
\begin{mathletters}
\label{3103}
\begin{eqnarray}
i{\bf \dot{\Psi}} &=&{\bf L\Psi ,}  \label{3103a} \\
{\bf L} &=&\left( 
\begin{array}{ccccccc}
\ddots  &  &  &  &  &  &  \\ 
\Omega _{+} & 2\Delta  & \Omega _{-} &  &  &  &  \\ 
& \Omega _{-} & \Delta  & \Omega _{+} &  &  &  \\ 
&  & \Omega _{+} & 0 & \Omega _{-} &  &  \\ 
&  &  & \Omega _{-} & -\Delta  & \Omega _{+} &  \\ 
&  &  &  & \Omega _{+} & -2\Delta  & \Omega _{-} \\ 
&  &  &  &  &  & \ddots 
\end{array}
\right) ,  \label{3103b}
\end{eqnarray}
where $\Omega _{\pm }=\Omega _{0}\cos \left( kz\pm \pi /4\right) ,\,$and $%
\Omega _{0}=g\sqrt{\nu }$ is a Rabi frequency. Let us denote by $\lambda
\left( z,t\right) $ the eigenvalue of the matrix ${\bf L}$ associated with a
dressed state (eigenvector of the matrix ${\bf L)}$ which coincides with the
initial state of the system before and after the interaction. As in the
cases above, if adiabaticity is maintained, the initial state wave function
undergoes a phase shift as a result of the interaction. The final state wave
function is still given by Eq. (\ref{11b}), with $\theta \left( z\right)
=\int_{-\infty }^{\infty }dt\lambda \left( z,t\right) .$

A remarkable situation had been found in reference \cite{c176}. With a Rabi
frequency $\Omega _{0}=\sqrt{3}\Delta /2,$ $\theta \left( z\right) $ closely
approximates a triangular potential. Explicitly, one finds 
\end{mathletters}
\begin{equation}
\theta \left( z\right) =-\theta w\left( z\right) ,  \label{3104}
\end{equation}
where $\theta =-\Delta \tau $ and the function $w\left( z\right) $ is shown
in Fig. \ref{f2}.

The slope of $w\left( z\right) $ has extrema at $z=0$ and $z=\lambda /4,$
where $w^{\prime }\approx \pm 0.624k,$ $w^{\prime \prime \prime }\approx \pm
0.211k^{3}.$ Using these values one finds that atoms scatter in two states
near $\tilde{n}_{0}=\pm n_{0},$ with 
\begin{equation}
n_{0}\approx 0.312\theta ,  \label{3105}
\end{equation}
The corresponding Fourier components are 
\begin{equation}
\psi _{n}\sim \left( \pm 1\right) ^{n}2.870\,n_{0}^{-1/3}Ai\left( \mp
2.870\,\Delta n/n_{0}^{1/3}\right) .  \label{3106}
\end{equation}
In this case the distribution $\left| \psi _{n}\right| ^{2}$ has width 
\begin{equation}
\Delta n_{bc}=0.57n_{0}^{1/3}.  \label{3107}
\end{equation}
This distribution is plotted in Fig. \ref{f1}. The distribution, obtained
here by a numerical integration of Eq. (\ref{13}) with pulse area (\ref{3104}%
), differs from the distribution calculated in the article \cite{Walls} by a
numerical solution of the Shr\"{o}dinger equation. We suppose that
differences arise owing to the fact that a rectangular pulse profile was
used in \cite{Walls}; such a pulse does not satisfy our adiabaticity
requirements.

\section{Atom grating profile}

The LABS split an initial $p=0$ state into two groups of states having
momenta close to $p=\pm 2n_{0}\hbar k.$ It is of interest to compare the
matter gratings generated by these LABS in the Fresnel scattering zone with
those of an ideal beam splitter. An ideal beam splitter produces purely
sinusoidal gratings having period 
\begin{equation}
d_{i}=\lambda /4n_{0}.  \label{311}
\end{equation}

At a time $t$ after scattering from a potential having period $\lambda /2,$
each momentum state's amplitude evolves as $\psi _{n}\exp \left( -i\omega
_{2k}n^{2}t\right) ,$ where $\omega _{2k}=2\hbar k^{2}/m$ is a recoil
frequency. The atom density $\rho \left( z,t\right) $ is given by 
\begin{equation}
\rho \left( z,t\right) =\sum_{n,n^{\prime }}\psi _{n}\psi _{n^{\prime
}}^{\ast }\exp \left[ 2ikz\left( n-n^{\prime }\right) -i\omega _{2k}t\left(
n^{2}-n^{\prime 2}\right) \right] .  \label{32}
\end{equation}

Asymptotic calculation of the atomic density profile is simplest for a
triangular potential (\ref{62}) (see Appendix). One finds in that case that
the grating profiles are not those of an ideal beam splitter. If one wishes
to carry out analogous solutions for the potentials considered in Sec. II,
purely numerical calculations are needed. Such calculations are not given in
this paper.

We are left with the problem of evaluating Eq. (\ref{32}) in some asymptotic
limit. Immediately we run into two problems: (1) Owing to the nonlinearities
in the potentials, one cannot simply use the asymptotic expression for the
Fourier components (\ref{16}) since, when these expressions are substituted
into Eq. (\ref{32}), one finds that the constraint, 
\begin{equation}
\sum \left| \psi _{n}\right| ^{2}=1,  \label{321}
\end{equation}
is violated. Equation (\ref{321}) follows from Eq. (\ref{32}) and the fact
that $2\int_{0}^{\lambda /2}\rho \left( z,t\right) dz/\lambda =1.$ If one
restricts the summation to values $n\approx \tilde{n}_{0},$ the series $\sum
\left| \psi _{n}\right| ^{2}$ calculated using Eq. (\ref{32}) diverges as $%
\sum \left( n-\tilde{n}_{0}\right) ^{-1/2}.$ This example demonstrates that
an alternative technique is needed to calculate the atomic density in the
asymptotic limit, which allows one to include contributions from all
momentum states. (2) Potentials having smooth rather than triangular extrema
lead to the atom focusing \cite{c5,c6} at a time $t\sim 1/\theta \omega
_{2k}$ \cite{c22}, i. e. to a sharp $\frac{\lambda }{2}$-period grating
having little in common with a sinusoidal grating. There is no simple way to
predict the times for which this density pattern is transformed into one
that might resemble that of the triangular potential. Rather than looking
for such times numerically, we consider an alternative approach for
obtaining the grating profile, valid for large times.

To this point we assumed implicitly that we were dealing with a monovelocity
atomic beam. The time $t$ is related to the observation plane at a distance $%
x$ from the interaction zone, by $t=x/u,$ where $u$ is the atom velocity
along the $x$-axis. In principle, for a LABS\ one needs a monovelocity beam,
since the pulse area $\theta $ and (therefore) the angle of scattering $%
\left( n_{0}\right) $ depends on $u$ like $1/u;$ however, one can neglect
this dependence for a small width of the velocity distribution, $\Delta u\ll
u.$ Although we require $\Delta u\ll u$, it is assumed that $\Delta u$ is
nonvanishing such that, for 
\begin{equation}
x\gg u^{2}/\Delta u\omega _{2k}  \label{33}
\end{equation}
the grating (\ref{32}) is an oscillating function of $u$ \cite{f}. On
averaging over $u$ one finds that the phase factor $\exp \left[ -i\omega
_{2k}t\left( n^{2}-n^{\prime 2}\right) \right] $ appearing in (\ref{32})
averages to zero {\em unless} \cite{c23} 
\begin{equation}
n^{\prime }=\pm n.  \label{34}
\end{equation}
Gratings of this type have been observed recently in the optical Talbot
effect \cite{c231}.

\begin{figure}
\begin{minipage}{0.97\linewidth}
\begin{center}
\epsfxsize=.5\linewidth \epsfysize=.63\linewidth \epsfbox{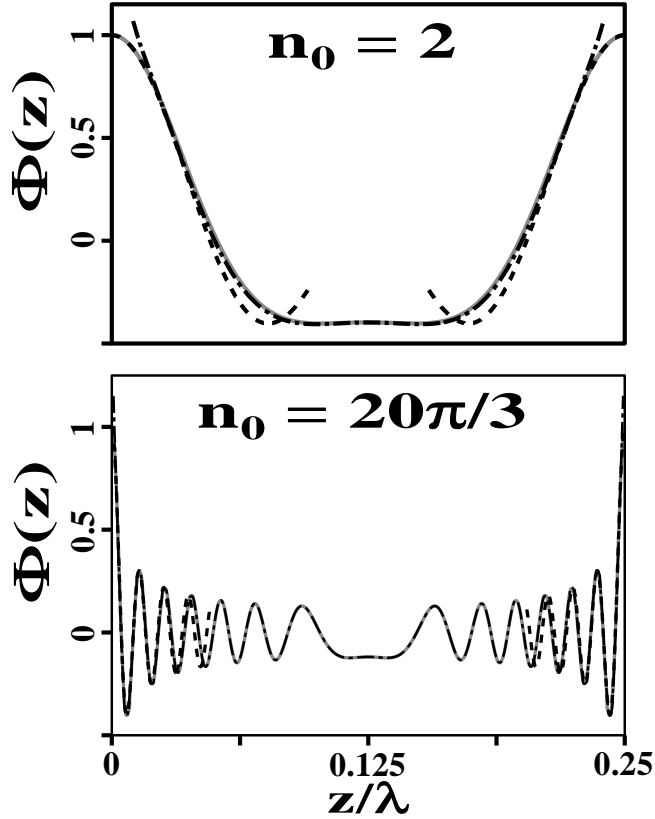}
\end{center}
\end{minipage}
\begin{minipage}{0.99\linewidth} \caption{Grating profile produced by a far-detuned standing wave field.
Solid gray lines are the exact profile calculated using Eqs. (\ref{36}, \ref
{37}), dashed and dot-dashed lines are asymptotic results (\ref{40a}, \ref
{40c}) and (\ref{40b}), respectively. 
\label{f3}}
\end{minipage}
\end{figure}

Setting $n^{\prime }=\pm n$ in Eq. (\ref{32}) one finds that the overall
grating period becomes $\lambda /4,$ instead of $\lambda /2.$ Inside a
period of the grating profile $\left( 0<z<\lambda /4\right) ,$ one arrives 
\cite{c23} at the following expression for the density: 
\begin{mathletters}
\label{35}
\begin{eqnarray}
\rho \left( z,t\right) &=&B+\Phi \left( z\right) ,  \label{35a} \\
B &=&\int_{0}^{\lambda /2}\frac{dz}{\lambda /2}\left| \eta \left( z\right)
\right| ^{2}-\left| \int_{0}^{\lambda /2}\frac{dz}{\lambda /2}\eta \left(
z\right) \right| ^{2}  \label{35b} \\
\Phi \left( z\right) &=&\frac{1}{2}\left[ F\left( z\right) +F\left(
z+\lambda /4\right) \right] ,  \label{35c} \\
F\left( z\right) &=&\int_{\left| \hat{z}\right| <Min\left\{ z,\frac{\lambda 
}{2}-z\right\} }\frac{d\hat{z}}{\lambda /4}\eta \left( z+\hat{z}\right) \eta
^{\ast }\left( z-\hat{z}\right) ,  \label{35d}
\end{eqnarray}
where $B$ is a constant background of order unity and $\eta \left( z\right) $
is the transmission function defined in Eq. (\ref{11a}). We will refer to $%
\Phi \left( z\right) $ as the {\em grating profile}. Although $\Phi \left(
z\right) $ can be negative, the atom density $\rho \left( z,t\right) \ $is
always positive$.$ For a phase transmission function (\ref{11b}) having
symmetric or antisymmetric pulse area $\left[ \theta \left( z\right) =\pm
\theta \left( -z\right) \right] ,$ one can simplify Eqs. (\ref{35c}, \ref
{35d}) as 
\end{mathletters}
\begin{mathletters}
\label{36}
\begin{eqnarray}
\Phi \left( z\right) &=&\frac{1}{2}\left[ F\left( z\right) +F\left( \lambda
/4-z\right) \right] ,  \label{36a} \\
F\left( z\right) &=&\int_{-z}^{z}\frac{d\hat{z}}{\lambda /4}\exp \left[
i\theta \chi \left( z,\hat{z}\right) \right] ,  \label{36b} \\
\chi \left( z,\hat{z}\right) &=&w\left( z+\hat{z}\right) -w\left( z-\hat{z}%
\right) ,  \label{36c}
\end{eqnarray}
where the functions $w\left( z\right) $ characterizing the various
potentials have been given in Sec. II.

\subsection{Standing wave field}

To simplify the calculations, we consider the asymptotic grating profile for
a far-detuned standing wave field $\left( \xi =2\chi ^{2}/\Delta ^{2}\ll
1\right) $ whose pulse area is given by Eq. (\ref{21a}). For this pulse $%
w(z)=\left\{ 1+\frac{\xi }{2}\left[ 1+\cos \left( 2kz\right) \right]
\right\} $ and 
\end{mathletters}
\begin{equation}
\chi \left( z,\hat{z}\right) =\xi \sin \left( 2kz\right) \sin \left( 2k\hat{z%
}\right)  \label{37}
\end{equation}
For large $n_{0}=\theta \xi /2\gg 1,$ one can evaluate the integral (\ref
{36b}) using a stationary phase method. The points of stationary phase are
at $\hat{z}=\pm \lambda /8.$ These points belong to the region of
integration if $z>\lambda /8.$ Thus, neglecting terms of the order of $%
1/\theta ,$ one finds 
\end{multicols}
\begin{equation}
F\left( z\right) \sim \Theta \left( z-\lambda /8\right) 2^{3/2}\left[ \pi
\theta \xi \sin \left( 2kz\right) \right] ^{-1/2}\cos \left[ \theta \xi \sin
\left( 2kz\right) -\pi /4\right] ,  \label{38}
\end{equation}
where $\Theta \left( z-\lambda /8\right) $ is a Heaviside step-function.
This expression becomes invalid near $z=\lambda /4$; however, for $\lambda
/4-z\lesssim \lambda /n_{0}$\ the integral (\ref{36b}) can be approximated
as 
\begin{equation}
F\left( z\right) \sim 2J_{0}\left[ \theta \xi \left( \pi -2kz\right) \right]
\label{39}
\end{equation}
Substituting Eqs. (\ref{38}, \ref{39}) in (\ref{36a}) one obtains 
\begin{mathletters}
\label{40}
\begin{eqnarray}
\Phi \left( z\right) &\sim &J_{0}\left( 4n_{0}kz\right) ,\text{ for }%
z\lesssim \lambda /n_{0};  \label{40a} \\
&\sim &\left[ \pi n_{0}\sin \left( 2kz\right) \right] ^{-1/2}\cos \left[
2n_{0}\sin \left( 2kz\right) -\pi /4\right] ,\text{ for }z\neq 0\text{ and }%
z\neq \lambda /4;  \label{40b} \\
&\sim &J_{0}\left[ 2n_{0}\left( \pi -2kz\right) \right] ,\text{ for }\lambda
/4-z\lesssim \lambda /n_{0}.  \label{40c}
\end{eqnarray}
The profile $\Phi \left( z\right) $ is shown in Fig. \ref{f3}. For large $%
n_{0}$ one sees from (\ref{40b}) that $\Phi \left( z\right) $ oscillates
rapidly with a spatially inhomogeneous period equal to 
\end{mathletters}
\begin{equation}
d\left( z\right) =d_{i}/\cos \left( 2kz\right) .  \label{42}
\end{equation}
Although there are rapid oscillations with reduced periodicity, the grating
does not approach the purity associated with an ideal beam splitter.

\subsection{Magneto-optical potential.}

For the magneto-optical scheme the pulse area, as a function of $z,$ has
discontinuous derivatives at multiples of $z=\lambda /8$ [see Eqs. (\ref{28}%
, \ref{29})]. Since both arguments $z\pm \hat{z}$ of the $w$-functions in
Eq. (\ref{36c}) vary in the interval $\left[ 0,2z\right] $ for $\left| \hat{z%
}\right| <z,$ one has to consider 4 cases: 
\begin{equation}
\alpha )\,\,z\in \left[ 0,\lambda /16\right] ,\,\,\beta )\,\,z\in \left[
\lambda /16,\lambda /8\right] ,\,\,\gamma )\,\,z\in \left[ \lambda
/8,3\lambda /16\right] ,\,\,\delta )\,\,z\in \left[ 3\lambda /16,\lambda /4%
\right] .  \label{43}
\end{equation}

For the simplest case $\alpha )$ one uses expression (\ref{29a}) for both
functions $w\left( z\pm \hat{z}\right) $ to find $\chi \left( z,\hat{z}%
\right) =2\cos \frac{2}{3}kz\sin \frac{2}{3}k\hat{z}.$ Stationary phase
points occur for $\hat{z}_{s}=\pm \frac{3}{8}\lambda \notin \left[ -z,z%
\right] $, and one can use an integration by parts in Eq. (\ref{36b}) to
arrive at the asymptotic expression 
\begin{equation}
F\left( z\right) \sim \left. 3\sin \left[ \theta \sin \left( \frac{4}{3}%
kz\right) \right] \right/ \pi \theta \cos ^{2}\left( \frac{2}{3}kz\right) .
\label{44}
\end{equation}
We will see below that only for small $z\sim 1/k\theta $ and for $\left|
z-\lambda /4\right| \sim 1/k\theta $ will terms of order $\theta ^{-1}$ be
important. In all other cases terms of order unity (arising for values of $z$
where $\chi \left( z,\hat{z}\right) \sim 1/\theta ),$ and terms of order $%
\theta ^{-1/2}$ (arising from integration near stationary phase points)
dominate. If such terms are absent, we say that corresponding contribution
vanishes.

\begin{figure}
\begin{minipage}{0.99\linewidth}
\begin{center}
\epsfxsize=.5\linewidth \epsfysize=.59\linewidth \epsfbox{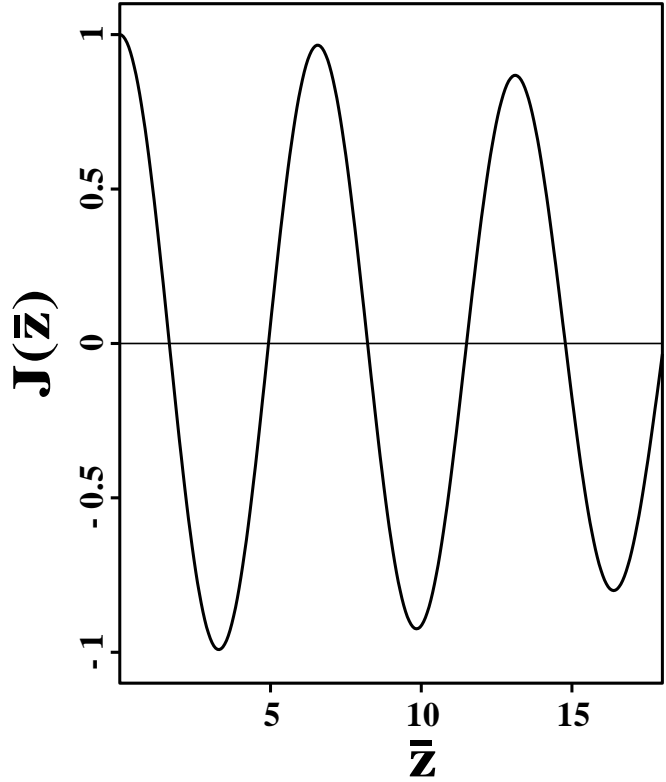}
\end{center}
\end{minipage}
\begin{minipage}{0.99\linewidth} \caption{$J(\bar{z})$ (\ref{48}) as a function of $\bar{z}$ near the points $%
z=\left( n+\frac{1}{2}\right) \frac{\protect\lambda }{4}.$ 
\label{f4}}
\end{minipage}
\end{figure}

In case $\beta )$ the function $w\left( z+\hat{z}\right) $ is equal to $\sin %
\left[ \frac{2}{3}k\left( z+\hat{z}\right) \right] $ and $\sin \left[ \frac{1%
}{3}\left( \pi -2k\left( z+\hat{z}\right) \right) \right] $ in the intervals 
$\hat{z}\in A_{+}=\left[ -z,\frac{\lambda }{8}-z\right] $ and $\hat{z}\in
B_{+}=\left[ \frac{\lambda }{8}-z,z\right] $ respectively, while the
function $w\left( z-\hat{z}\right) $ is equal to $\sin \left[ \frac{2}{3}%
k\left( z-\hat{z}\right) \right] $ and $\sin \left[ \frac{1}{3}\left( \pi
-2k\left( z-\hat{z}\right) \right) \right] $ in the intervals $\hat{z}\in
A_{-}=\left[ z-\frac{\lambda }{8},z\right] $ and $\hat{z}\in B_{-}=\left[
-z,z-\frac{\lambda }{8}\right] .$ Since $B_{\mp }\subset A_{\pm }$ one has
to map out the region of integration $\left[ -z,z\right] $ into 3 intervals: 
\begin{equation}
a=\left[ -z,z-\frac{\lambda }{8}\right] ,\,\,b=\left[ z-\frac{\lambda }{8},%
\frac{\lambda }{8}-z\right] ,\,\,c=\left[ \frac{\lambda }{8}-z,z\right] .
\label{45}
\end{equation}
Let us denote by $F_{i}\left( z\right) $ the contribution to $F\left(
z\right) $ from interval $i$ $\left( i=a,\,\,b,\,\,c\right) $ For $\hat{z}%
\in a,$ $\chi \left( z,\hat{z}\right) =2\sin \left[ \frac{1}{6}\left(
4kz-\pi \right) \right] \cos \left[ \frac{1}{6}\left( \pi +4k\hat{z}\right) %
\right] ,$ $\hat{z}_{s}=-\lambda /8\notin a,$ but the function $\chi $
becomes small near $z=\lambda /8,$ i. e. 
\begin{mathletters}
\label{46}
\begin{eqnarray}
F_{a}\left( z\right) &\sim &\frac{3}{\pi }\int_{0}^{\pi /6}d\phi \exp \left[
-\frac{4}{3}i\theta k\left( \frac{\lambda }{8}-z\right) \cos \phi \right] ,%
\text{ for }\frac{\lambda }{8}-z\lesssim 1/k\theta ,  \label{46a} \\
&\sim &0,\text{ otherwise.}  \label{46b}
\end{eqnarray}
The functions $F_{b}\left( z\right) =0$, $F_{c}\left( z\right) =F_{a}^{\ast
}\left( z\right) ,$ and\ one finds for case $\beta )$%
\end{mathletters}
\begin{mathletters}
\label{47}
\begin{eqnarray}
F\left( z\right) &\sim &J\left[ \frac{4}{3}\theta k\left( \frac{\lambda }{8}%
-z\right) \right] ,\text{ for }\frac{\lambda }{8}-z\lesssim 1/\theta ,
\label{47a} \\
&\sim &0,\text{ otherwise},  \label{47b}
\end{eqnarray}
where 
\end{mathletters}
\begin{equation}
J\left( \bar{z}\right) =\frac{6}{\pi }\int_{0}^{\pi /6}d\phi \cos \left[ 
\bar{z}\cos \phi \right] .  \label{48}
\end{equation}
Function (\ref{48}) is plotted in Fig. \ref{f4}.

In case $\gamma )$ one has to distinguish intervals $a=\left[ -z,\frac{%
\lambda }{8}-z\right] ,$ $b=\left[ \frac{\lambda }{8}-z,z-\frac{\lambda }{8}%
\right] ,$ $c=\left[ z-\frac{\lambda }{8},z\right] .$ When $\hat{z}\in a,$ $%
\chi \left( z,\hat{z}\right) =2\sin \left[ \frac{1}{6}\left( 4kz-\pi \right) %
\right] \cos \left[ \frac{1}{6}\left( 4k\hat{z}+\pi \right) \right] ,$ $\hat{%
z}_{s}=-\lambda /8\in a$ and the leading contribution to $F_{a}\left(
z\right) $ is 
\begin{equation}
F_{a}\left( z\right) \sim 3\left\{ \pi \theta \sin \left[ \frac{1}{6}\left(
4kz-\pi \right) \right] \right\} ^{-1/2}\exp \left\{ 2i\theta \sin \left[ 
\frac{1}{6}\left( 4kz-\pi \right) \right] -i\frac{\pi }{4}\right\} ,
\label{49}
\end{equation}
while $F_{b}\left( z\right) =0$, $F_{c}\left( z\right) =F_{a}^{\ast }\left(
z\right) .$ Adding these contributions one finds for case $\gamma )$%
\begin{mathletters}
\label{50}
\begin{eqnarray}
F\left( z\right) &\sim &2K\left( z\right) ,  \label{50a} \\
K\left( z\right) &=&3\left\{ \pi \theta \left| \sin \left[ \frac{1}{6}\left(
4kz-\pi \right) \right] \right| \right\} ^{-1/2}\cos \left\{ 2\theta \left|
\sin \left[ \frac{1}{6}\left( 4kz-\pi \right) \right] \right| -\frac{\pi }{4}%
\right\}  \label{50b}
\end{eqnarray}
In the region near $\lambda /8$ $\left[ 0<z-\lambda /8\lesssim 1/\theta %
\right] ,$ where Eq. (\ref{50}) is invalid, one returns back to expression (%
\ref{47a}).

Another feature arises in case $\delta ),$ when one has to map out the
integration region $\hat{z}\in \left[ -z,z\right] $ into 5 intervals: 
\end{mathletters}
\begin{equation}
a=\left[ -z,z-\frac{3}{8}\lambda \right] ,\,\,b=\left[ z-\frac{3}{8}\lambda ,%
\frac{\lambda }{8}-z\right] ,\,\,c=\left[ \frac{\lambda }{8}-z,z-\frac{%
\lambda }{8}\right] ,\,\,d=\left[ z-\frac{\lambda }{8},\frac{3}{8}\lambda -z%
\right] ,\,\,e=\left[ \frac{3}{8}\lambda -z,z\right]  \label{51}
\end{equation}
The stationary phase point $z_{s}$ belongs to the region of integration for
the contribution $F_{b}\left( z\right) ,$ where $\chi \left( z,\hat{z}%
\right) =2\sin \left[ \frac{1}{6}\left( 4kz-\pi \right) \right] \cos \left[ 
\frac{1}{6}\left( 4k\hat{z}+\pi \right) \right] ,$ $z_{s}=-\lambda /8$ and $%
F_{b}\left( z\right) $ is given by the right-hand-side of Eq. (\ref{49}) .
To achieve this result one needs the region of integration to be larger than 
$\lambda /\theta ^{1/2},$ but the interval $b$ in Eq. (\ref{51}) has zero
length at $z=\lambda /4,$ rendering the method of stationary phase invalid.
In this region\ the function $F_{b}\left( z\right) $ is simply equal to the
integrand at $\hat{z}=z_{s}$ multiplied by the length of interval $b:$%
\begin{equation}
F_{b}\left( z\right) \approx \frac{4}{\pi }k\delta z\exp \left[ i\theta
\left( 1-2\cdot 3^{-1/2}k\delta z\right) \right] ,\text{ for }z=\lambda
/4-\delta z,\,\,\delta z\lesssim \lambda /\theta .  \label{52}
\end{equation}
Calculating all other contributions using integration by parts, one arrives
at the following final result for region near $z$=$\lambda /4$ 
\begin{equation}
F\left( z\right) =\frac{4}{\pi \theta }\left\{ 2k\delta z\theta \cos \left[
\theta \left( 1-2\cdot 3^{-1/2}k\delta z\right) \right] +3^{1/2}\sin \left[
\theta \left( 1-2\cdot 3^{-1/2}k\delta z\right) \right] -\frac{3}{4}\sin
\left( \frac{4}{3}\theta k\delta z\right) \right\} .  \label{53}
\end{equation}
Substituting $F\left( z\right) $ into Eq. (\ref{36a}) one finds $\left[
0\leq z\leq \lambda /8\right] $%
\begin{mathletters}
\label{54}
\begin{eqnarray}
\Phi \left( z\right) &\sim &\frac{2}{\pi \theta }\left\{ 2kz\theta \cos %
\left[ \theta \left( 1-2\cdot 3^{-1/2}kz\right) \right] +3^{1/2}\sin \left[
\theta \left( 1-2\cdot 3^{-1/2}kz\right) \right] \right\} ,\text{ for }%
z\lesssim \lambda /\theta ;  \label{54a} \\
\Phi \left( z\right) &\sim &J\left[ \frac{4}{3}\theta k\delta z\right] ,%
\text{ for }z=\frac{\lambda }{8}-\delta z,\,\,\delta z\lesssim \lambda
/\theta ;  \label{54b} \\
\Phi \left( z\right) &\sim &K\left( z\right) \text{ otherwise},  \label{54c}
\end{eqnarray}
where the functions $J\left( x\right) $ and $K\left( z\right) $ are given by
Eqs. (\ref{48}, \ref{50b}) respectively. The profile for $\lambda
/8<z<\lambda /4$ can be obtained using the symmetry property, $\Phi \left( 
\frac{\lambda }{4}-z\right) =\Phi \left( z\right) .$ The profile (\ref{54})
is shown in Fig. \ref{f5}.

\begin{figure}
\begin{minipage}{0.99\linewidth}
\begin{center}
\epsfxsize=.5\linewidth \epsfysize=.7\linewidth \epsfbox{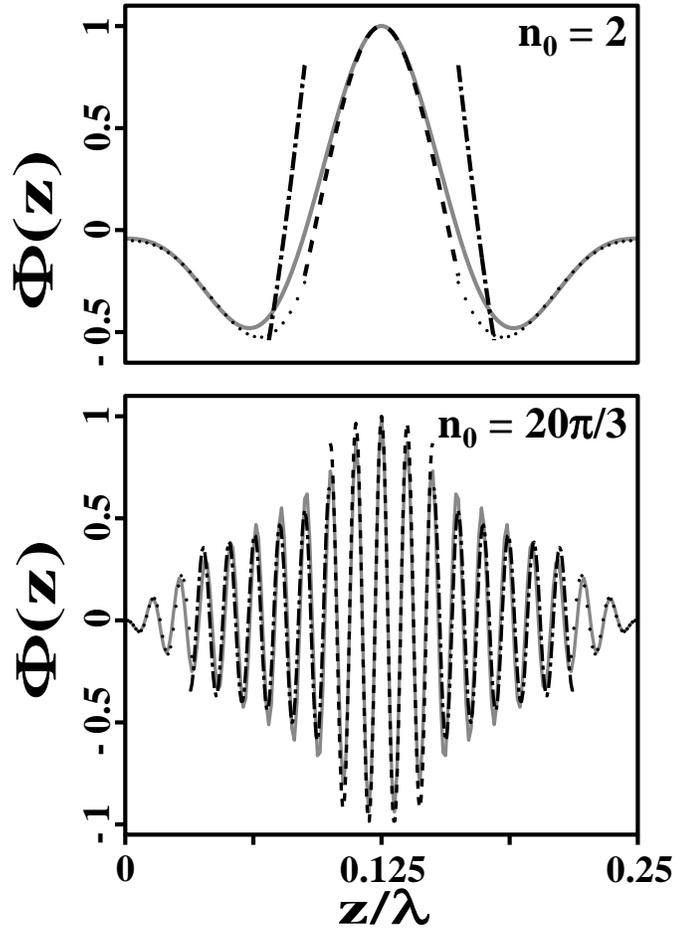}
\end{center}
\end{minipage}
\begin{minipage}{0.99\linewidth} \caption{The same as Fig \ref{f3}, but for scattering by a magneto-optical
field. Dashed, dot-dashed and dotted lines are asymptotic dependences (\ref
{54b}, \ref{54c}, \ref{54a}) respectively. 
\label{f5}}
\end{minipage}
\end{figure}

As in the case of the standing wave beam splitter, everywhere except near
the points $z=0,$ $\lambda /8,$ $\lambda /4,$ the profile is a grating
having a spatially inhomogeneous period 
\end{mathletters}
\begin{equation}
d\left( z\right) =d_{i}\left/ \cos \left[ \frac{1}{6}\left( 4kz-\pi \right) %
\right] \right. .  \label{55}
\end{equation}
The profile is not that of an ideal beam splitter.

\begin{multicols}{2}
\subsection{Bichromatic beam splitter}

\begin{figure}
\begin{minipage}{0.99\linewidth}
\begin{center}
\epsfxsize=.97\linewidth \epsfysize=1.121\linewidth \epsfbox{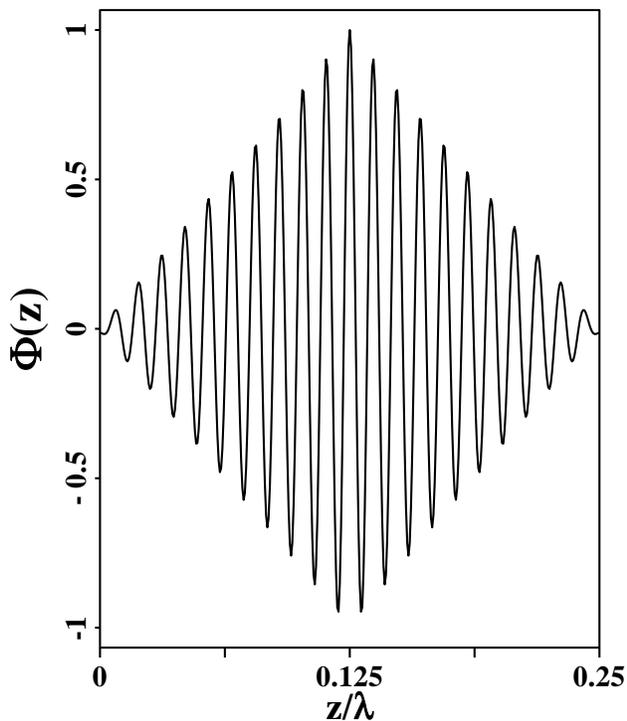}
\end{center}
\end{minipage}
\begin{minipage}{0.99\linewidth} \caption{Density profile produced by bichromatic field for $n_{0}=20\protect%
\pi /3.$ 
\label{f6}}
\end{minipage}
\end{figure}

Lacking an analytical expression for the potential produced by a bichromatic
field we are not able to obtain an analytical asymptotic expression for the
grating profile. On the other hand, it is possible to evaluate Eqs. (\ref
{36a}) numerically. In Fig. \ref{f6} the profile is shown for $n_{0}=20\pi
/3 $ and is similar to that for the magneto-optical potential.

\section{Discussion}

Asymptotic expressions in the limit of large-angle scattering have been
obtained for the atomic wave function and grating profile after scattering
from a periodic triangular potential (see Appendix) and from potentials
which approximate this triangular potential. The potentials are chosen \cite
{c13,c176} with the goal of producing two and {\em only} two final state
momenta 
\begin{equation}
p=\pm 2n_{0}\hbar k,  \label{69}
\end{equation}
with $\left( n_{0}\gg 1\right) .$ In turn, interference between these
momenta components in the Fresnel scattering region lead to a pure,
sinusoidal high-order grating 
\begin{equation}
\rho \left( z\right) =1+\cos \left( 4n_{0}kz\right)   \label{70}
\end{equation}
having period 
\begin{equation}
d_{i}=\lambda /4n_{0}.  \label{71}
\end{equation}

In practice, it is impossible to produce only two final state momenta with
periodic potentials. At best, one finds scattering into two {\em groups }of
states centered around $p=\pm 2n_{0}\hbar k$. Owing to the range of momenta
in each group, none of the potentials can generate the ideal grating profile
(\ref{70}). For the potentials considered in this work, the width of the
momentum groups scales asymptotically as $n_{0}^{1/3}$ [see Eqs. (\ref{19}, 
\ref{31}, \ref{3107})]. If one's goal is to creates a higher order
sinusoidal grating, then one needs to increase the angle of splitting, i. e.
the parameter $n_{0}.$ However, by increasing $n_{0}$ one broadens the width
of the momentum groups, and diminishes degree to which the grating profile
approximates a pure sinusoidal grating.

In the limit of strong fields and large scattering angles, we developed
asymptotic techniques for evaluating the atomic wave function and density.
One might think that a quasiclassical or WKB method could be used in this
limit, but the linear nature of the potential invalidates such an approach
in the region of maximum scattering. Instead, the method of stationary phase
was used to obtain asymptotic expressions for the wave function. The results
depend solely on the first and third derivatives of the potential at the
point of maximum slope. For all but the triangular potential, however, these
asymptotic expressions cannot be used in the expressions for the atomic
density since they do not produce a convergent result.

Only for the triangular potential does the asymptotic expression for the
Fourier amplitudes converge rapidly enough as a function of Fourier index $n$
to allow one to calculate the atomic density using these asymptotic
formulas. The periodic triangular potential is analyzed in the Appendix. The
scattered signal is a periodic function of time with period $t_{0}=T/2n_{0}$%
, where $T=2\pi /\omega _{2k}$ is the Talbot time (recall that time $t$ is
related to the distance $x$ following the atom-field interaction zone by $%
t=x/u$, where $u$ is the longitudinal velocity). In a given scattering plane
following the atom-field interaction, the atom density consists of three
distinct spatial regions, one of the form (\ref{70}) but with twice as large
an amplitude, one with $\rho \left( z\right) =1,$ and one with $\rho \left(
z\right) =0$. Examples of these profiles are shown in Fig. \ref{f7} in the
Appendix.

For the standing wave, magneto-optical, and bichromatic field potentials,
the atomic density was calculated for scattering distances $x\gg
Tu^{2}/\Delta u$ \cite{c24}, where $\Delta u\ll u$ is the range of
longitudinal velocities\cite{c23}. The density is a periodic function of $z$
having period $\lambda /4.$

For the standing wave potential one can distinguish regions about $%
z=n\lambda /4$ ($n$ is an arbitrary integer) of extent $\lambda /n_{0}$
where the density has a sharp maximum [see Eq. (\ref{40a})] of order unity.
With the exception of these regions the density is an oscillating function
of $z$ with amplitude of order $n_{0}^{-1/2}$ and {\em spatially
inhomogeneous} period (\ref{42}). The period becomes infinite at $z=$ $%
\left( n+\frac{1}{2}\right) \frac{\lambda }{4},$ which implies that, near
these points, the profile is flat. All these features are seen in Fig. \ref
{f3}, where the grating profile is plotted for $n_{0}=2$ and $n_{0}=20\pi
/3. $ The last case corresponds to the value chosen in \cite{c15} for
numerical calculations. It is surprising that the analysis of the scattering
in terms of two distinct spatial intervals seems to be valid even for $%
n_{0}=2,$ when the condition $n_{0}\gg 1$ is not satisfied.

For the magneto-optical case one has to distinguish three regions 
\begin{equation}
\begin{tabular}{l}
$\text{I }\left| z-\left( n+\frac{1}{2}\right) \frac{\lambda }{4}\right|
\lesssim \lambda /n_{0,}$ \\ 
$\text{II. }\left| z-n\frac{\lambda }{4}\right| \lesssim \lambda /n_{0,}$ \\ 
III. other values of $z.$%
\end{tabular}
\label{72}
\end{equation}
In region I the modulated component of the grating profile $\Phi (z)$ is
described by the function $J\left[ \frac{4}{3}k\theta \left( \frac{\lambda }{%
8}-z\right) \right] $ [see Eqs. (\ref{54b}, \ref{48})]; in region I this
function is of order unity and oscillates with a slowly decaying amplitude
(see Fig. \ref{f2}). In region II the grating profile oscillates with an
amplitude of order $n_{0}^{-1}$. In the remaining region III, which accounts
for most of the range when $n_{0}\gg 1$, the grating profile is similar to
that found for the standing wave potential, i. e. it is a grating having
amplitude of order $n_{0}^{-1/2}$ and a spatially inhomogeneous amplitude
period given by Eq. (\ref{55}). In contrast to the standing wave case, this
period cannot be larger than $2\times 3^{-1/2}d_{i}.$ The grating profile
produced by the magneto-optical field is shown in the Fig. \ref{f4}. For a
moderate value $n_{0}=2,$ the dependence (\ref{54c}) associated with region
III is a poor approximation to the profile; however, one can see from Fig. 
\ref{f4} that it is not necessary to use the expression for region III.
Regions I and II overlap with region III and with each other, and Eqs. (\ref
{54b}, \ref{54a}) provide a good approximation to the profile for all $z.$

The grating profile for the bichromatic potential is shown in Fig. \ref{f6}
for $n_{0}=20\pi /3.$

In summary, we have asymptotic expressions for the momentum distribution and
calculated the grating profile for several classes of optical beam
splitters. While the momentum distribution resulting from these beam
splitters can approach that of an ideal beam splitter, the associated
grating profiles do not closely approximate a pure sinusoidal pattern.

\acknowledgments

This work is supported by the U. S. Army Research Office under Grant No.
DAAD19-00--1-0412 and by the National Science Foundation under Grant No.
PHY-9800981. We are grateful to J. L. Cohen, G. Raithel, Y. V.
Rozhdestvensky, T. Sleator, L. P. Yatsenko for fruitful discussions.
\end{multicols}

\appendix

\section{Appendix - Triangular Potential.}

It is instructive to consider scattering by the ideal periodic triangular
potential (\ref{62}). This potential differs from those considered in the
text in that the second derivative in each half-period of the potential
vanishes identically. The pulse area associated with this potential is given
by 
\begin{equation}
\theta \left( z\right) =2kn_{0}\left\{ 
\begin{tabular}{l}
-$z,$ for $\left| z\right| <\lambda /8$ \\ 
$z-\lambda /4,$ for $\left| z-\lambda /4\right| <\lambda /8$%
\end{tabular}
\right. ,  \label{56}
\end{equation}
where $n_{0}=\bar{n}/2.$ \ For the Fourier components (\ref{13}), one finds 
\begin{equation}
\psi _{n}=\left\{ 
\begin{tabular}{l}
$2n_{0}\sin \left[ \left( n-n_{0}\right) \pi /2\right] \left/ \pi \left(
n^{2}-n_{0}^{2}\right) \right. ,$ for $n\neq -n_{0}$ \\ 
$\left( -1\right) ^{n_{0}}/2,$ for $n=-n_{0}$%
\end{tabular}
\right. .  \label{57}
\end{equation}

\begin{figure}
\begin{minipage}{0.99\linewidth}
\begin{center}
\epsfxsize=.7\linewidth \epsfysize=.61\linewidth \epsfbox{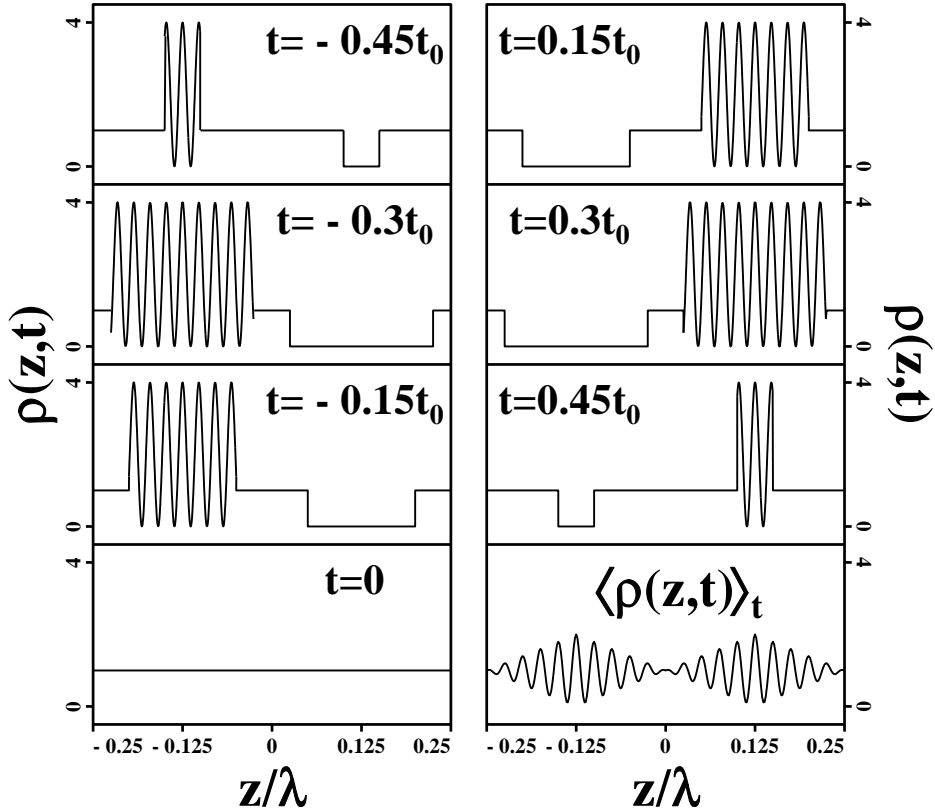}
\end{center}
\end{minipage}
\begin{minipage}{0.99\linewidth} \caption{Scattering by the triangular potential. The first seven graphs give
the density profile for a single period at different times. The last graph
if the density averaged over $t_{0}$. The time $t_{0}$ is the Talbot time
divided by $n_{0}$ and $n_{0}=11$ in this figure. 
\label{f7}}
\end{minipage}
\end{figure}

Consider now the case of large integer $n_{0}.$ For $n$ close to $\pm n_{0}$
one finds 
\begin{mathletters}
\label{59}
\begin{eqnarray}
\psi _{n_{0}+\Delta n} &\approx &\sin \left[ \Delta n\pi /2\right] \left/
\pi \Delta n\right. ,  \label{59a} \\
\psi _{-n_{0}+\Delta n} &\approx &(-1)^{n_{0}}\left[ \delta \left( \Delta
n\right) -\psi _{n_{0}+\Delta n}\right] .  \label{59b}
\end{eqnarray}
where $\delta \left( \Delta n\right) $ is a Kronecker symbol. \ In a
contrast to the nonlinear potentials considered above, the amplitudes (\ref
{59}) decay sufficiently fast to calculate both the atomic wave function and
atom density. In the expression for the wave function, 
\end{mathletters}
\begin{mathletters}
\label{60}
\begin{eqnarray}
\psi \left( z,t\right)  &=&\psi _{+}\left( z,t\right) +\psi _{-}\left(
z,t\right) ,  \label{60a} \\
\psi _{\pm }\left( z,t\right)  &=&\sum_{\Delta n}\psi _{\pm n_{0}+\Delta
n}\exp \left[ 2ikz\left( \pm n_{0}+\Delta n\right) -i\omega _{2k}\left( \pm
n_{0}+\Delta n\right) ^{2}t\right]   \label{60b}
\end{eqnarray}
one can neglect terms quadratic in $\Delta n$ in the exponents if $\left|
t-mT\right| \ll T$, where $m$ is an integer and $T=2\pi /\omega _{2k}$ is
the Talbot time. For these values of $t$, one can obtain analytic asymptotic
expressions for the wave function and density. The wave function is a
periodic function of $z$ having period $\lambda /2$ and a periodic function
of $t$ having a {\em reduced} Talbot period 
\end{mathletters}
\begin{equation}
t_{0}=\pi /\omega _{2k}n_{0}.  \label{63}
\end{equation}
Within one period 
\begin{mathletters}
\label{64}
\begin{eqnarray}
\psi _{+}\left( z,t\right)  &=&\exp \left( 2in_{0}kz-i\omega
_{2kn_{0}}t\right) \Theta \left( \frac{\lambda }{8}-\left| z-2n_{0}\frac{%
\hbar k}{m}t\right| \right) ,  \label{64a} \\
\psi _{-}\left( z,t\right)  &=&\exp \left( -2in_{0}kz-i\omega
_{2kn_{0}}t\right) \Theta \left( \left| z+2n_{0}\frac{\hbar k}{m}t\right| -%
\frac{\lambda }{8}\right) ,  \label{64b}
\end{eqnarray}
where $\Theta \left( x\right) $ is a Heaviside step-function.

One sees that after scattering the wave function splits in two components
having average momentum $\pm 2n_{0}\hbar k.$ Moreover, there is a spatial
modulation of the wave function corresponding to $\frac{\lambda }{4}$
sections of the incident plane wave scattered with momenta $\pm 2n_{0}\frac{%
\hbar k}{m}$. These wave function components, having unit amplitude and
spatial width $\frac{\lambda }{4}$, are centered at $z=z_{\pm },$ where 
\end{mathletters}
\begin{mathletters}
\label{64}
\begin{eqnarray}
z_{+} &=&2n_{0}\frac{\hbar k}{m}t+n_{+}\frac{\lambda }{2},  \label{65a} \\
z_{-} &=&-2n_{0}\frac{\hbar k}{m}t+\left( n_{-}+\frac{1}{2}\right) \frac{%
\lambda }{2}  \label{65b}
\end{eqnarray}
($n_{\pm }$ are arbitrary integers). Interference occurs when different
components centered at $z_{\pm }$ overlap. Outside of the interference
regions, the density can equal unity for points belonging to one of the $%
z_{\pm }$-regions or $0$ for other points. Analyzing all these
possibilities, one arrives at the following density profile 
\end{mathletters}
\begin{mathletters}
\label{66}
\begin{eqnarray}
\rho \left( z,t\right) &=&\xi \left( z,t\right) +2\left( -1\right)
^{n_{0}}\cos \left( 4n_{0}kz\right) \phi \left( z,t\right) ,  \label{66a} \\
\phi \left( z,t\right) &=&\left\{ 
\begin{array}{ll}
1 & 
\begin{array}{l}
\text{if }z\in \left[ -\frac{3\lambda }{8}-z_{t},\frac{\lambda }{8}+z_{t}%
\right] \text{ and }t\in \left[ -\frac{t_{0}}{2},-\frac{t_{0}}{4}\right] \\ 
\text{if }z\in \left[ -\frac{\lambda }{8}+z_{t},-\frac{\lambda }{8}-z_{t}%
\right] \text{ and }t\in \left[ -\frac{t_{0}}{4},0\right] \\ 
\text{if }z\in \left[ \frac{\lambda }{8}-z_{t},\frac{\lambda }{8}+z_{t}%
\right] \text{ and }t\in \left[ 0,\frac{t_{0}}{4}\right] \\ 
\text{if }z\in \left[ -\frac{\lambda }{8}+z_{t},\frac{3\lambda }{8}-z_{t}%
\right] \text{ and }t\in \left[ \frac{t_{0}}{4},\frac{t_{0}}{2}\right]
\end{array}
\\ 
0 & \text{in other cases}
\end{array}
\right. ;  \label{66b} \\
\xi \left( z,t\right) &=&\left\{ 
\begin{array}{l}
0,\text{ if }z\in \left[ -\frac{\lambda }{8}-z_{t},\frac{3\lambda }{8}+z_{t}%
\right] \\ 
1,\text{ if }z\in \left[ -\frac{\lambda }{4},-\frac{3\lambda }{8}-z_{t}%
\right] \cup \left[ \frac{\lambda }{8}+z_{t},-\frac{\lambda }{8}-z_{t}\right]
\cup \left[ \frac{3\lambda }{8}+z_{t},\frac{\lambda }{4}\right] \\ 
2,\text{ if }z\in \left[ -\frac{3\lambda }{8}-z_{t},\frac{\lambda }{8}+z_{t}%
\right]
\end{array}
\right\} ,\text{ for }t\in \left[ -\frac{t_{0}}{2},-\frac{t_{0}}{4}\right] ;
\label{66c} \\
&=&\left\{ 
\begin{array}{l}
0,\text{ if }z\in \left[ \frac{\lambda }{8}+z_{t},\frac{\lambda }{8}-z_{t}%
\right] \\ 
1,\text{ if }z\in \left[ -\frac{\lambda }{4},-\frac{\lambda }{8}+z_{t}\right]
\cup \left[ -\frac{\lambda }{8}-z_{t},\frac{\lambda }{8}+z_{t}\right] \cup 
\left[ \frac{\lambda }{8}-z_{t},\frac{\lambda }{4}\right] \\ 
2,\text{ if }z\in \left[ -\frac{\lambda }{8}+z_{t},-\frac{\lambda }{8}-z_{t}%
\right]
\end{array}
\right\} ,\text{ for }t\in \left[ -\frac{t_{0}}{4},0\right] ;  \label{66d} \\
&=&\left\{ 
\begin{array}{l}
0,\text{ if }z\in \left[ -\frac{\lambda }{8}-z_{t},-\frac{\lambda }{8}+z_{t}%
\right] \\ 
1,\text{ if }z\in \left[ -\frac{\lambda }{4},-\frac{\lambda }{8}-z_{t}\right]
\cup \left[ -\frac{\lambda }{8}+z_{t},\frac{\lambda }{8}-z_{t}\right] \cup 
\left[ \frac{\lambda }{8}+z_{t},\frac{\lambda }{4}\right] \\ 
2,\text{ if }z\in \left[ \frac{\lambda }{8}-z_{t},\frac{\lambda }{8}+z_{t}%
\right]
\end{array}
\right\} ,\text{ for }t\in \left[ 0,\frac{t_{0}}{4}\right] ;  \label{66e} \\
&=&\left\{ 
\begin{array}{l}
0,\text{ if }z\in \left[ -\frac{3\lambda }{8}+z_{t},\frac{\lambda }{8}-z_{t}%
\right] \\ 
1,\text{ if }z\in \left[ -\frac{\lambda }{4},-\frac{3\lambda }{8}+z_{t}%
\right] \cup \left[ \frac{\lambda }{8}-z_{t},-\frac{\lambda }{8}+z_{t}\right]
\cup \left[ \frac{3\lambda }{8}-z_{t},\frac{\lambda }{4}\right] \\ 
2,\text{ if }z\in \left[ -\frac{\lambda }{8}+z_{t},\frac{3\lambda }{8}-z_{t}%
\right]
\end{array}
\right\} ,\text{ for }t\in \left[ \frac{t_{0}}{4},\frac{t_{0}}{2}\right] ;
\label{66f}
\end{eqnarray}
where 
\end{mathletters}
\[
z_{t}=\frac{\lambda t}{2t_{0}}. 
\]
This profile is shown in Fig. \ref{f7}. The function $\phi \left( z,t\right) 
$ represents an envelope function for higher order gratings having period $%
d_{i}=\lambda /4n_{0}.$

Expression (\ref{66b}) gives the envelope at a given time. One can consider
also the grating profile as a function of $t$ for given $z$. This would
correspond to atoms arriving at a given $z$ with different longitudinal
velocities, allowing one to make a comparison with the results of Sec. III.
For fixed $z$ as a function of $t$, one finds 
\begin{equation}
\phi \left( z,t\right) =\left\{ 
\begin{array}{ll}
1 & 
\begin{array}{l}
\text{if }t\in \left[ -\frac{3}{4}t_{0}-2\frac{z}{\lambda }t_{0},\frac{1}{4}%
t_{0}+2\frac{z}{\lambda }t_{0}\right] \text{ and }z\in \left[ -\frac{\lambda 
}{4},-\frac{\lambda }{8}\right] \\ 
\text{if }t\in \left[ -\frac{1}{4}t_{0}+2\frac{z}{\lambda }t_{0},-\frac{1}{4}%
t_{0}-2\frac{z}{\lambda }t_{0}\right] \text{ and }z\in \left[ -\frac{\lambda 
}{8},0\right] \\ 
\text{if }t\in \left[ \frac{1}{4}t_{0}-2\frac{z}{\lambda }t_{0},\frac{1}{4}%
t_{0}+2\frac{z}{\lambda }t_{0}\right] \text{ and }z\in \left[ 0,\frac{%
\lambda }{8}\right] \\ 
\text{if }t\in \left[ -\frac{1}{4}t_{0}+2\frac{z}{\lambda }t_{0},\frac{3}{4}%
t_{0}-2\frac{z}{\lambda }t_{0}\right] \text{ and }z\in \left[ \frac{\lambda 
}{8},\frac{\lambda }{4}\right]
\end{array}
\\ 
0 & \text{in other cases}
\end{array}
\right.  \label{67}
\end{equation}
One sees that over one period of Talbot oscillations the high harmonic
amplitude is constant for some times and vanishes for others. The harmonic
amplitude averaged over a Talbot oscillation period is simply given by $%
\left\langle \phi \left( z,t\right) \right\rangle _{t}=t_{z}/t_{0},$ where $%
t_{z}$ is the duration of the time interval when the harmonic amplitude is
constant; the background term on average is equal to unity $\left[
\left\langle \xi \left( z,t\right) \right\rangle _{t}=1\right] .$
Calculating $t_{z}$ from \ref{67}$,$ one finds 
\begin{eqnarray}
\left\langle \rho \left( z,t\right) \right\rangle _{t} &=&1+2\left(
-1\right) ^{n_{0}}\cos \left( 4n_{0}kz\right)   \nonumber \\
&&\times \left\{ 
\begin{array}{l}
1-\frac{4}{\lambda }\left| z\right| ,\text{ for }\left| z\right| >\frac{%
\lambda }{8} \\ 
\frac{4}{\lambda }\left| z\right| ,\text{ for }\left| z\right| <\frac{%
\lambda }{8}
\end{array}
\right\} .  \label{68}
\end{eqnarray}
This result could be obtained using Eqs. (\ref{35}, \ref{36}) of Sec. III
with the pulse area (\ref{56}).

\begin{multicols}{2}

\end{multicols}


\begin{references}
\bibitem{c1}  B. Dubetsky, A. P. Kazantsev, V. P. Chebotayev, V. P.
Yakovlev, Pis'ma Zh. Eksp. Teor. Fiz. {\bf 39}, 531 (1984) [JETP Lett. {\bf %
39}, 649 (1985)].

\bibitem{c2}  A. P. Kazantsev, G. I. Surdutovich, V. P. Yakovlev, Pis'ma Zh.
Eksp. Teor. Fiz. {\bf 31}, 542 (1980) [JETP Lett. {\bf 31}, 509 (1980)].

\bibitem{c3}  P. E. Moskowitz, P. L.Gould, S. R. Atlas, and D. E. Pritchard,
Phys. Rev. Lett. {\bf 51},370,1983.

\bibitem{c4}  E. M. Rasel, M. K. Oberthaler, H. Batelaan, J. Schmiedmayer,
A. Zeilinger, Phys. Rev. Lett. {\bf 75}, 2633 (1995).

\bibitem{c5}  M. Prentiss, G. Timp, N. Bigelow, R. E. Behringer, J. E.
Cunningham, Appl. Phys. Lett. {\bf 60}, 1027, (1992).

\bibitem{c6}  T. Sleator, T. Pfau, V. Balykin, J. Mlynek, Appl. Phys. B {\bf %
54}, 375, (1992).

\bibitem{c8}  B. Dubetsky and P. R. Berman, Phys. Rev. A {\bf 50}, 4057
(1994).

\bibitem{c9}  P. R. Berman, B. Dubetsky, and J. L. Cohen, Phys. Rev. A {\bf %
58}, 4801 (1998).

\bibitem{c10}  D. M. Giltner, R. W. McGowan, S. A. Lee, Phys. Rev. Lett. 
{\bf 75}, 2638 (1995).

\bibitem{J}  J. L. Cohen, Thesis, University of Michigan, 2000, pp. 140-143.

\bibitem{c11}  A. V. Turlapov, D. V. Strekalov, A. Kumarakrishnan, S. Cahn,
and T. Sleator, in {\it ICONO 98: Quantum Optics, Interference Phenomena in
Atomic Systems, and High-Precision Measurements} (edited by A. V. Andreev,
S. N. Bagayev, A. S. Chirkin, and V. I. Denisov), SPIE Proc. {\bf 3736},
26-37 (1999).

\bibitem{c12}  P. R. Berman and B. Bian, Phys. Rev. A {\bf 55}, 4382 (1997).

\bibitem{c13}  R. Grimm, V. S. Letokhov, Yu. B. Ovchinnikov, A. I. Sidorov,
J. Phys. II France, {\bf 2}, 593 (1992).

\bibitem{c14}  T. Pfau, Ch. Kurtsiefer, C. S. Adams, M. Sigel, J. Mlynek,
Phys. Rev. Lett. {\bf 71}, 3427 (1993).

\bibitem{c15}  C. S. Adams, T. Pfau, Ch. Kurtsiefer, J. Mlynek, Phys. Rev.
A, {\bf 48}, 2108 (1993).

\bibitem{c16}  U. Janicke, M. Wilkens, Phys. Rev. A, {\bf 50}, 3265 (1994).

\bibitem{p3}  A. P. Chu, K. S. Johnson, and M. G. Prentiss, J. Opt. Soc. Am.
B {\bf 13}, 1352 (1996).

\bibitem{c17}  J. Soding, R. Grimm, Phys. Rev. A, {\bf 50}, 2517 (1994).

\bibitem{p1}  K. S. Johnson, A. Chu, T. W. Lynn, K. K. Berggren, M. S.
Shahriar, and M. Prentiss, Opt. Lett. {\bf 20}, 1310 (1995).

\bibitem{c171}  V. S. Voitsekhovich, M. V. Danileiko, A. M. Negriiko, V. I.
Romanenko, and L. P. Yatsenko, Zh. Tekh. Fiz. {\bf 58}, 1174 (1988) [Sov.
Phys. Tech. Phys. {\bf 33,} 690 (1988)].

\bibitem{c172}  V. S. Voitsekhovich, M. V. Danileiko, A. M. Negriiko, V. I
Romanenko, and L. P. Yatsenko, Pis'ma Zh. Eksp.Teor. Fiz. {\bf 49}, 138
(1989) [JETP Lett. {\bf 49,} 161 (1989)].

\bibitem{c173}  R. Grimm, Yu. B. Ovchinnikov, A. I. Sidorov, V. S. Letokhov,
Phys. Rev. Lett. {\bf 65}, 1415 (1990);

\bibitem{c174}  J. Soding, R. Grimm, Yu. B. Ovchinnikov, P. Bouyer, and Ch.
Salomon, Phys. Rev. Lett. {\bf 78}, 1420 (1997).

\bibitem{c175}  M. Williams, F. Chi, M. Cashen, and H. Metcalf, Phys. Rev. 
{\bf A 61}, 023408 (2000).

\bibitem{c176}  R. Grimm, J. Soding, Yu. B. Ovchinnikov, Opt. Lett. 19, 658
(1994).

\bibitem{p2}  K. S. Johnson, A. P. Chu, K. K. Berggren, and M. G. Prentiss,
Optics Comm. {\bf 126, }326 (1996).

\bibitem{Walls}  S. M. Tan, D. F. Walls, Opt. Commun. {\bf 118}, 412 (1995).

\bibitem{c177}  T. Wong, M. K. Olsen, S. M. Tan, D. F. Walls, Phys. Rev. A 
{\bf 52}, 2161 (1995).

\bibitem{c178}  E A Korsunsky, Yu. B. Ovchinnikov, Optics Comm. {\bf 143},
219 (1997).

\bibitem{c179}  A. S. Pazgalev and Yu. V. Rozhdestvenski, Pisma Zh. Eksp.
Teor. Fiz. {\bf 66}, 386 (1997) [JETP Lett., {\bf 66}, 412 (1997)].

\bibitem{c180}  E. A. Korsunsky, Quantum Semiclass. Opt. {\bf 10}, 477
(1998).

\bibitem{c181}  V. I. Romanenko and L. P. Yatsenko, Zh. Eksp. Teor. Fiz. 
{\bf 117}, 467 (2000).[JETP {\bf 90}, 407 (2000)].

\bibitem{c18}  P. R. Berman, Phys. Rev. A {\bf 53}, 2627 (1996).

\bibitem{G}  P. L. Gould, D. E. Pritchard, In Proceedings of the
International School of Physics ''Enrico Fermi'', Course CXXXI, edited by A.
Aspect, W. Barletta and R. Bonifacio (Amsterdam, Oxford, Tokio, Washington
DC) 1996, p. 443.

\bibitem{c19}  M. Abramowitz and I. A. Stegun editors, ''{\it Handbook of
Mathematical Functions},'' Dover Publications, Inc. (New York, 1972), p.
366, No. 9.3.4.

\bibitem{c22}  J. L. Cohen, B. Dubetsky, and P. R. Berman, Phys. Rev. A {\bf %
60}, 4886 (1999).

\bibitem{f}  The distance $x$ is always confined to the Fresnel scattering
zone, which for a beam of diameter $b$ requires $x\phi \ll b,$ where $\phi
\sim n_{0}\hbar k/Mu$ is the scattering angle.

\bibitem{c23}  B. Dubetsky and P. R. Berman, In ''{\it Atom Interferometry}%
'', edited by P. R. Berman (Academic Press, Cambridge, MA, 1997), p 407 (or
http://arXiv.org/PS\_cache/physics/pdf/0005/0005078. pdf), Sec. VI.

\bibitem{c231}  N. Guerineau, B. Harchaoui, and J. Primot, Opt. Commun. {\bf %
180}, 199 (2000).

\bibitem{c24}  This condition could be too strong. For example in the
triangular potential the averaged profile arises for distances $x\gg
t_{0}u^{2}/\Delta u$ , which are $2n_{0}$ times smaller than those given by
Eq. (\ref{33}).
\end{references}
\end{document}